\documentstyle[preprint,aps,epsfig]{revtex}

\begin{document}

\title{The $b\to s\gamma\gamma$ transition in softly broken supersymmetry}

\author{S. Bertolini\\
INFN, Sezione di Trieste\\[-1ex]
Scuola Internazionale Superiore di Studi Avanzati\\[-1ex]
via Beirut 4, I-34013 Trieste, Italy.}
\author{J. Matias\thanks{Address from December 1997: CERN, Th-Division,
CH-1211 Gen\`eve 23, Switzerland}\\[-1ex]
Dipartimento di Fisica, Universit\`a di Padova\\[-1ex]
via F. Marzolo 8, I-35131 Padova, Italy.}

\maketitle

\begin{abstract}

We study the effect of supersymmetric contributions to the
effective quark transition $b\to s\gamma\gamma$, including leading
order QCD effects. We apply the discussion to the decay
$B_s\to\gamma\gamma$. Even though one-particle
irreducible contributions could play a role, numerical cancelations
make the
amplitude for the two-photon emission strongly correlated to the
$b\to s\gamma$ amplitude which is sharply constrained by experiment.
A quite general statement follows: as long as non standard physics
effects appear only in the matching of the Wilson coefficients of the standard
effective operator basis, the deviations
from the standard model expectations
of the decay rates induced by $b\to s\gamma\gamma$
are bound to follow closely the corresponding deviations on
$b\to s\gamma$. Effects of new physics are therefore bound
to be small.

\end{abstract}


\begin{flushright}
\begin{tabular}{l}
{ SISSA 107/97/EP \hspace{1cm} ~~~} \\[-1ex]
{ September 1997    ~~~}
\end{tabular}
\end{flushright}

\newpage


\section{\bf Introduction}

The rare flavour changing transition $b\to s\gamma\gamma$ has recently
attracted new interest in view of the planned experiments at the upcoming
KEK and SLAC B-factories and existing hadronic accelerators, which may test
branching fractions as low as $10^{-8}$ times the B meson decay width.

Rare B-physics potentially provides us with valuable redundacy tests of the
flavour structure of the standard model (SM) and complementary
information on the related CP violation.
It is also most sensitive to the ``heavy'' sector of the SM particle
spectrum and it is the preferred low-energy laboratory
for virtual signals of exotic physics.

In recent years the $b\to s\gamma$ decay has provided us with the first
experimental evidence of ``penguin'' physics and the experimental bounds
on the $B\to X_s \gamma$ decays have shown to provide sharp constraints
on the physics beyond the SM.

The fact that the $b\to s\gamma$ transition has been already experimentally
observed is related to the peculiar enhancement of the electroweak
amplitude that arises at the two-loop level due to a large logarithmic
QCD mixing with the effective $b\to s \bar c c$ operator\cite{first,lo}.
The study of the QCD leading logarithmic (LO) resummation for $b\to s\gamma$
(and $b\to s\ gluon$) has been very recently extended to the next-to-leading
order~\cite{prenlo,nlo,postnlo}
thus reducing the theoretical estimated error for the
inclusive rate below the 10\% threshold.

In order to produce similar studies for the $b\to s\gamma\gamma$ transition
it is extremely helpful to observe that, by the use of an extension
of Low's low energy theorem~\cite{low,yao1} or, alternatively,
by applying the equation
of motions, the $b\to s\gamma\gamma$ quark operator can be expanded at
$O(G_f)$ on the standard operator basis needed for $b\to s \gamma$.

Three groups have recently presented a QCD LO calculation of the
two photon transition~\cite{hill2,yao2,reina}
 thus improving on the previous electroweak
calculations~\cite{yao1,simma1,herrlich}.

More recently a study of the $b\to s\gamma\gamma$ decay in two-Higgs
doublet extensions of the SM has appeared~\cite{aliev}.

The purpose of the present paper is to study the influence on the
$b\to s\gamma\gamma$ transition of softly broken supersymmetry.
We find actually the results of our analysis having a more general impact
on the possible effects of new physics for the radiative two-photon
decay. At the LO the short-distance part of the
$b\to s\gamma\gamma$ amplitude
turns out to be controlled by the one-photon radiative component.
The higher dimension one-particle irreducible contributions present
in the $b\to s\gamma\gamma$ amplitude, 
potentially large because of the $1/m_c^2$ dependence, turn out
to remain subleading because of accidental cancelations.
This result remains true when studying two-Higgs
doublet extensions of the SM 
where the additional charged Higgs contribution to the
$b\to s\gamma$ component of the two-photon amplitude adds coherently to
the SM one. On the other hand,
in supersymmetric theories there are potentially large destructive
effects related to the exchange of superpartners, which may
reduce the size of the one-particle reducible part of the
$b\to s\gamma\gamma$ amplitude thus affecting the relative weight
between the latter and the one-particle irreducible components.

Nevertheless, the present
experimental constraints on $b\to s\gamma$ induced decays are tight, 
and by studying
the effects of low energy supersymmetry as a paradigmatic case, 
the present analysis shows generally that,
as long as extensions of the SM affect only the short distance
Wilson coefficients of the standard effective Hamiltonian,
the $b\to s\gamma\gamma$ induced decays are severely constrained by
the present bounds on the inclusive $B\to X_s \gamma$ decay~\cite{CLEO}
\begin{equation}
BR(B\to X_s \gamma)=(2.32\pm 0.51\pm 0.29\pm 0.32)\times 10^{-4}.
\end{equation}
Since this result is consistent within 30\% with the next-to-leading
order (NLO) SM calculation~\cite{nlo}
\begin{equation}
BR(B\to X_s \gamma)=(3.28\pm 0.33)\times 10^{-4},
\end{equation}
one may not expect much larger deviations of the two-photon decay rates
from the corresponding SM estimates.
Were one to invoke supersymmetry to reduce by 30\%
the SM prediction of $\Gamma(B\to X_s \gamma)$ then the SUSY rates related
to $b\to s\gamma\gamma$ will unlikely
exceed the corresponding SM expectations. Potential
implications of a (albeit challenging) NLO analysis of the
two-photon amplitude in extensions of the SM are commented
upon in the conclusions.

Although we shall apply our results to the $B_s\to\gamma\gamma$ decay
we are not interested in predicting absolute decay rates
and thus we will not discuss the uncertainties related to
hadronization, which are not affected by short-distance new physics.
Our main purpose is to study the deviations of the quark
amplitudes from the SM predictions.

On the basis of these considerations and at present time,
it is not worthy the effort
to perform a detailed analysis of a specific supersymmetric model.
We will investigate the main features of the SUSY amplitudes
by means of relatively simple limiting cases in the supersymmetric
and soft breaking parameter space.

\section{\bf Effective Quark Lagrangian: Operator Basis and
Coefficients}

 At the LO in QCD the effective
Hamiltonian for $b\to s \gamma$ closes on a basis of eight
effective operators
\begin{equation}
H_{eff} =-{G_F\over
\sqrt{2}}V^*_{ts}V_{tb}\sum_{i=1}^{8}C_i(\mu)O_i(\mu),
\label{heff}
\end{equation}
where
\begin{eqnarray}
O_1 &=& (\bar{s}_ic_j)_{V-A}(\bar{c}_jb_i)_{V-A} \nonumber \\
O_2 &=& (\bar{s}_ic_i)_{V-A}(\bar{c}_jb_j)_{V-A}  \nonumber \\
O_3 &=& (\bar{s}_ib_i)_{V-A}\sum_{q}(\bar{q}_jq_j)_{V-A}  \nonumber
\\
O_4 &=& (\bar{s}_ib_j)_{V-A}\sum_{q}(\bar{q}_jq_i)_{V-A}  \nonumber
\\
O_5 &=& (\bar{s}_ib_i)_{V-A}\sum_{q}(\bar{q}_jq_j)_{V+A}  \nonumber
\\
O_6 &=& (\bar{s}_ib_j)_{V-A}\sum_{q}(\bar{q}_jq_i)_{V+A}  \nonumber
\\
O_7 &=& {e\over
8\pi^2}\bar{s}_i\sigma^{\mu\nu}(m_s(1-\gamma_5)
                               +m_b(1+\gamma_5))b_iF_{\mu\nu} \nonumber
\\
O_8 &=& {g\over
8\pi^2}\bar{s}_i\sigma^{\mu\nu}(m_s(1-\gamma_5)
                               +m_b(1+\gamma_5))T^a_{ij}b_jG^a_{\mu\nu}\ .
\label{operators}
\end{eqnarray}
In eq.~(\ref{operators})
$i,j=1,2,3$ are color indices,
$a=1,...,8$ labels the $SU(3)$ generators,
and $V\pm A\equiv 1\pm\gamma_5$.
The sum runs over the active quark flavours $u,d,s,c,b$.

Having factored out the relevant Kobayashi-Maskawa (KM) mixings,
the LO matching of the Wilson coefficients $C_i(\mu)$ at the
scale $m_W$ is given in the SM by
\begin{equation}
C^{SM}_i(m_W) = 0,\ \ \ \ \ i=1,3,4,5,6
\label{c1-6}
\end{equation}
\begin{equation}
C^{SM}_2(m_W) = 1
\label{c2}
\end{equation}
\begin{equation}
C^{SM}_7(m_W) = \frac{3 x^3-2 x^2}{4(x-1)^4}\log x - \frac{8 x^3 + 5 x^2 -
                    7 x}{24(x-1)^3}
\label{c7}
\end{equation}
\begin{equation}
C^{SM}_8(m_W) = \frac{-3 x^2}{4(x-1)^4}\log x - \frac{x^3 - 5 x^2 - 2
x}{8(x-1)^3}
\label{c8}
\end{equation}
where $x=m_t^2/m_W^2$.

Large $(\alpha_s\log\mu)^n$ corrections to the weak
amplitudes are resummed via the
renormalization group (RG) equations by evolving the Wilson
coefficients at the typical scale of the process ($\mu\approx m_b$).
While the anomalous dimension matrix of the first six operators
is regularization scheme independent,
the operator mixings between
$O_1,\cdots, O_6$, and the dipole penguins $O_{7,8}$
are generally regularization-scheme dependent since they arise in LO
at the two-loop level. However,
the finite one-loop matrix elements
of $b\to s\gamma$ (and $b\to s g$) generated by the insertion
of the $O_{5,6}$ gluonic penguins
carry a compensating
scheme-dependence such that the total physical amplitudes are
independent on the regularization
scheme. In view of this, one defines the so called  ``effective''
Wilson coefficients~\cite{ciuc,buras},
$C^{eff}_7(\mu)$ and $C^{eff}_8(\mu)$, for which the LO RG
running is scheme independent.
As an example, for $C_7(\mu)$ one finds the LO expression
\begin{equation}
C^{eff}_7(\mu) = \eta^{\frac{16}{23}} C_7(m_W) +
\frac{8}{3} \left( \eta^{\frac{14}{23}} - \eta^{\frac{16}{23}}
\right) C_8(m_W)  + C_2(m_W) \sum_{i=1}^8 h_i \eta^{a_i}
\label{c7eff}
\end{equation}
where $\eta=\alpha_s(m_W)/\alpha_s(\mu)$ and
the numbers $h_i$ and $a_i$ are given in the appendix
of ref.~\cite{buras}. The sum of all $h_i$ is zero and
$C_i^{eff}(m_W)=C_i(m_W)$.
For notational convenience,
the superscript ``eff'' on  $C_{7,8}(\mu)$ will be henceforth
omitted.

\section{\bf The $B_s\to\gamma\gamma$ Decay}

The quark $b\to s\gamma\gamma$ transition induces at the 
hadronic level two interesting rare decay modes of the $B$ mesons:
$B_{u,d}\to X_s\ \gamma\gamma$ and $B_s\to\gamma\gamma$,
where $X_s$ represents strange mesonic states.
Both these decay modes and their features at the
hadronic level within the SM have been widely studied in 
the literature~\cite{bxsgg}.

We apply the analysis of the QCD corrected supersymmetric
$b\to s\gamma\gamma$ amplitude
to the discussion of the two body $B_s\to\gamma\gamma$ decay, 
whose total rate can be cast in a simple and compact form; this allows us
to carry out the present analysis without unnecessary complications. 
Our conclusions are based on short-distance
properties of the quark transition that hold as well for the
$B_{u,d}\to X_s\ \gamma\gamma$ decay, even though the detailed form 
of the amplitude differs from that of the $B_s\to\gamma\gamma$ decay. 

The total $B_s\to\gamma\gamma$
amplitude can be separated into a CP-even and a CP-odd part
\begin{equation}
{\cal A}(B_s\to \gamma\gamma)=M^+F_{\mu\nu}F^{\mu\nu}
+iM^-F_{\mu\nu}\tilde{F}^{\mu\nu}.
\end{equation}
According to the notation of ref.~\cite{yao2} one finds
\begin{equation}
M^+=-{4{\sqrt 2}\alpha_{em} G_F\over
9\pi}f_{B_s}m_{B_s}V_{ts}^*V_{tb}\left( B(\mu)\ m_b\ K(m_b^2)
+{3C_7(\mu)\over 8\bar{\Lambda} }\right),
\end{equation}
and
\begin{equation}
M^-={4{\sqrt 2}\alpha_{em} G_F\over
9\pi}f_{B_s}m_{B_s}V_{ts}^*V_{tb}\left(\sum_q m_{B_s}\
A_q(\mu)\ J(m_q^2)+m_b\ B(\mu)\ L(m_b^2)+
{3C_7(\mu)\over 8\bar{\Lambda} }\right),
\end{equation}
where $\bar{\Lambda}=m_{B_s}-m_b$ and
\begin{eqnarray}
A_u &=&(C_3-C_5)N_c+(C_4-C_6)\nonumber \\
A_d &=&{1\over 4}\left((C_3-C_5)N_c+(C_4-C_6)\right)\nonumber \\
A_c &=&(C_1+C_3-C_5)N_c+(C_2+C_4-C_6) \nonumber \\
A_s &=&A_b={1\over 4}\left((C_3+C_4-C_5)N_c+(C_3+C_4-C_6)\right)
\nonumber \\
B &=&-{1\over 4}(C_6N_c+C_5)\ ,
\label{OPI}
\end{eqnarray}
are combinations of Wilson coefficients evaluated at the scale
$\mu$ ($\approx m_b$).

The functions $J(m^2)$, $K(m^2)$ and $L(m^2)$  are defined by
\begin{eqnarray}
J(m^2)&=&I_{11}(m^2),\nonumber \\
K(m^2)&=&4\ I_{11}(m^2)-I_{00}(m^2)\ ,\nonumber \\
L(m^2)&=&I_{00}(m^2),
\end{eqnarray}
with
\begin{equation}
I_{pq}(m^2)=\int_{0}^{1}{dx}\int_{0}^{1-x}{dy}{x^py^q\over
m^2-2xyk_1\cdot k_2-i\varepsilon}\ ,
\label{Ipq}
\end{equation}
and $2\ k_1\cdot k_2=m_{B_s}^2$.

The decay width for $B_s\to \gamma\gamma$ is then given by
\begin{equation}
\Gamma(B_s\to \gamma\gamma)={m_{B_s}^3\over 16\pi}({\vert M^+\vert
}^2+{\vert M^-\vert }^2)\ ,
\end{equation}
and,
from the measured $B_s$ lifetime,
the corresponding branching ratio is finally obtained
\begin{equation}
BR(B_s\to \gamma\gamma)=\Gamma(B_s\to \gamma\gamma)/\Gamma(B_s)\ .
\label{brff}
\end{equation}

Coming to the inclusive $B_s\to X_s \gamma$ decay it is convenient to use
the approximate equality
\begin{equation}
BR(B\to X_s \gamma) \simeq
\frac{\Gamma(b\to s \gamma)}{\Gamma(b\to c e\bar\nu_e)}
BR(B\to X_c e\bar\nu_e)\ ,
\label{brsf}
\end{equation}
where
\begin{equation}                   \label{main}
\frac{\Gamma(b \to s \gamma)}{\Gamma(b \to c e \bar{\nu}_e)}
 =  \frac{|V_{ts}^* V_{tb}|^2}{|V_{cb}|^2}
\frac{6 \alpha_{em}}{\pi g(z)} |C_7(\mu)|^2\ ,
\label{rquark}
\end{equation}
which minimizes the uncertainties related to the bottom quark mass
and KM mixings.
In eq.~(\ref{rquark}) the function
$g(z) = 1 - 8z^2 + 8z^6 - z^8 - 24z^4 \log z$
is the phase space factor in the semileptonic decay
and $z = m_c/m_b$.

In Figs. 1 and 2
we show the LO results for the SM $B_s\to \gamma\gamma$
and $B\to X_s \gamma$ branching ratios, as a function
of the renormalization scale $\mu$ and of $m_t$.
Our numerical results are obtained using the values given in
Table~\ref{inputs} for the other input parameters.

\epsfxsize=8cm
\centerline{\epsfbox{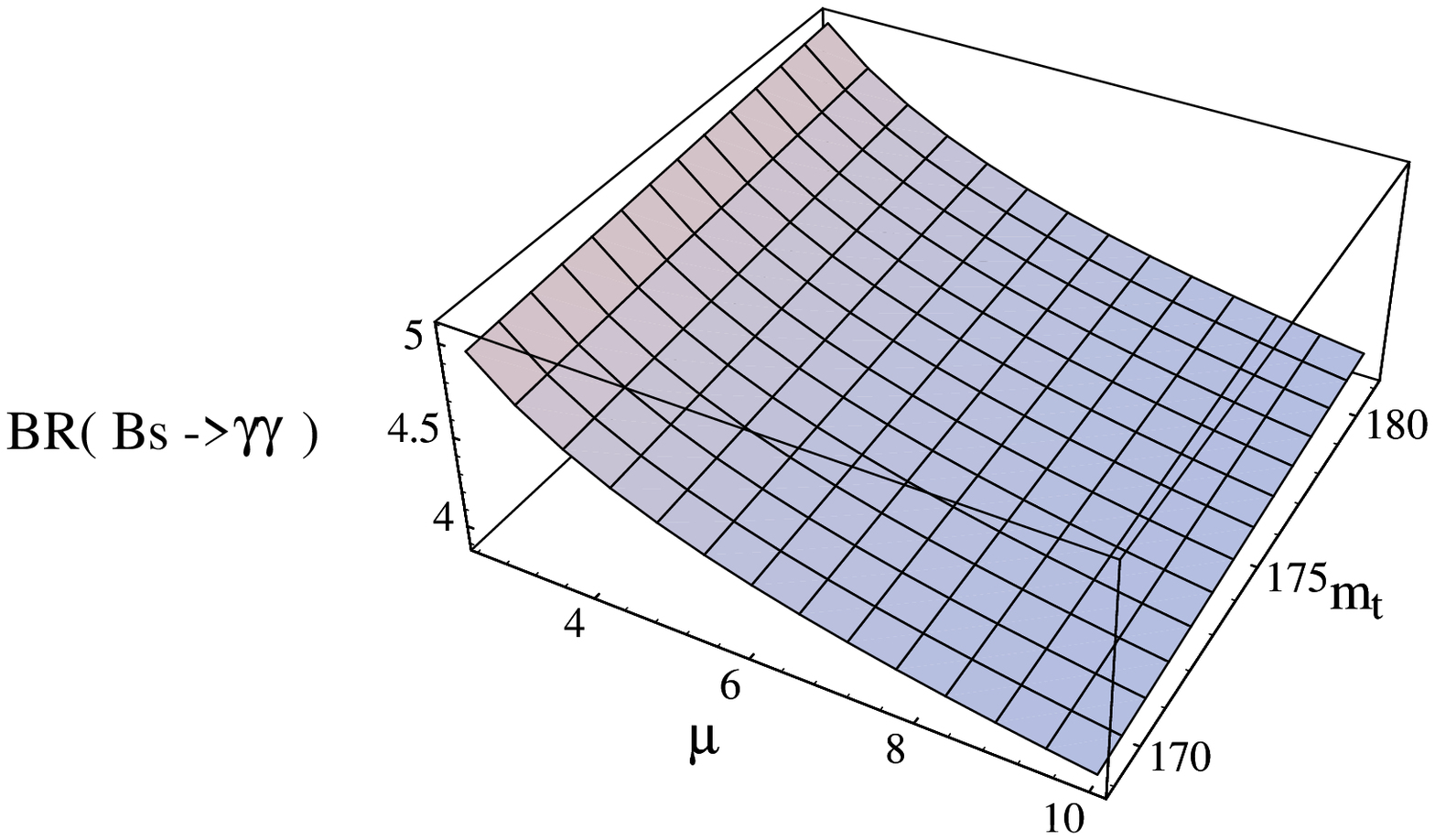}}
\label{mfig1}
{Fig. 1.
$BR(B_s\to\gamma\gamma)_{SM} \times 10^{7}$
as a function of $\mu$ and $m_t$ (GeV)
for central values of the other input parameters (Table I).
}

\epsfxsize=8cm
\centerline{\epsfbox{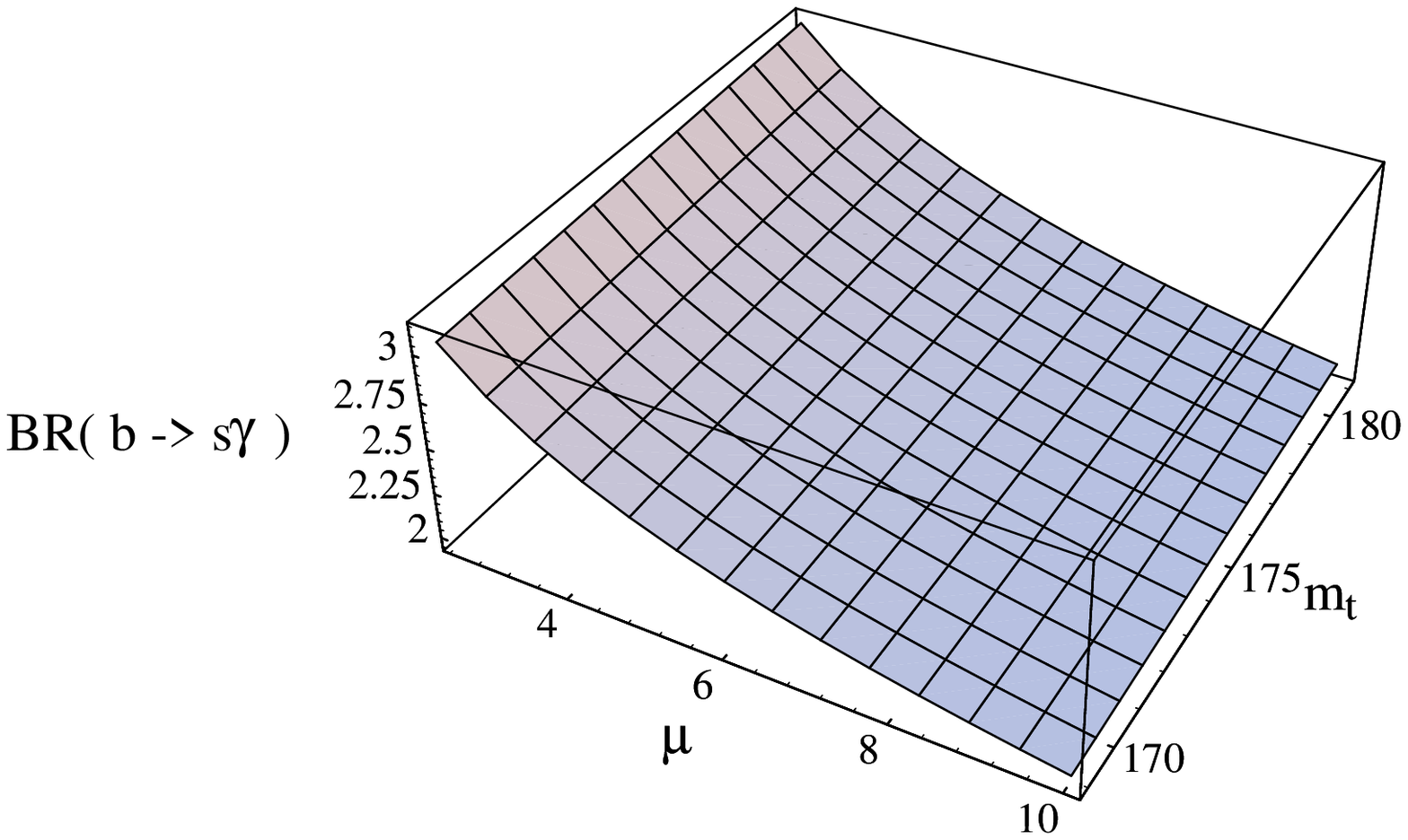}}
\label{mfig2}
{Fig. 2.
$BR(B\to X_s \gamma)_{SM} \times 10^{4}$
as a function of $\mu$ and $m_t$ (GeV)
for central values of the other input parameters (Table I).
}
\bigskip

From a direct comparison of Figs. 1 and 2 it appears clear that the
$B_s\to\gamma\gamma$ decay rate is dominated by the $C_7$
component. This is related to the fact that
the one-particle irreducible contributions arising from the operators
$Q_{1,2}$, which could be potentially large due to
the $1/m_c^2$ dependence of eq. (\ref{Ipq}), appear in eq. (\ref{OPI}) 
($A_c$) via the combination $N_c\ C_1 + C_2$ which is numerically 
suppressed~\cite{yao2}. 
Notice that the scale dependence represents the largest
source of uncertainty of the LO calculation~\cite{buras,ali}.
We have here shown the range 2.5 GeV $< \mu <$ 10 GeV.

As a reference for the following analysis,
the SM central values for the LO QCD corrected decay rates
at the scale $\mu=m_b$ are given by
\begin{equation}
BR(B\to X_s \gamma)_{SM} = 2.5\times 10^{-4}
\label{brsfsm}
\end{equation}
and
\begin{equation}
BR(B_s\to \gamma\gamma)_{SM} = 4.4\times 10^{-7}\ .
\label{brffsm}
\end{equation}
The prediction in eq.~(\ref{brffsm}) compares to the present
experimental bound~\cite{L3}
\begin{equation}
BR(B_s\to\gamma\gamma)<1.48\times 10^{-4}\
\label{ffexp}
\end{equation}
which is about three orders of magnitude away from the needed sensitivity.

\section{\bf The $B_s\to\gamma\gamma$ Decay in Softly Broken
Supersymmetry}

In a wide class of realistic SUSY models
the global supersymmetry
breaking is a consequence of the spontaneous breaking
of an underlying N=1 supergravity theory
(for reviews see ref. \cite{SUGRA}).
The locally supersymmetric lagrangian is supposed to undergo a spontaneous
breaking in the so called hidden sector,
and the effects of this breaking are communicated to the
observable sector through gravitational effects.
A renormalizable theory is obtained in
the limit in which the Planck mass goes to infinity. By doing so
one is left with an effective
globally supersymmetric lagrangian and explicit soft breaking terms.
In our present study we shall consider the following gauge invariant
soft breaking Lagrangian:
\begin{equation}
{\cal L}_{soft}= -{\cal M}^2-(\hat {M}+S\ +\ h.c.)
\label{hslagr}
\end{equation}
where
${\cal M}^2$ is a common mass term for all the scalar components
$z_i$ in the theory
\begin{equation}
{\cal M}^2 \equiv  \Sigma_i \tilde m^2 z_i^* z_i\ ,
\label{soft}
\end{equation}
$\hat {M}$ is a mass term for the gauginos $\lambda_\alpha,$
$\alpha=1,2,3$ considered as Weyl fields
\begin{equation}
\hat{M} \equiv -\frac{M_\alpha}{2} \lambda_\alpha \lambda_\alpha\ ,
\label{gm}
\end{equation}
and
$S$ is the scalar analogue of the superpotential
\begin{equation}
S = \tilde m  \left[ -A_U h_U H_2 \widetilde{Q} \widetilde{U}^c
+ A_D h_D H_1 \widetilde{Q} \widetilde{D}^c
+ A_E h_E H_1 \tilde{L} \widetilde{E}^c   +
B  \mu H_1 H_2 \right]\ ,
\label{trilbi}
\end{equation}
where with standard notation
$h_{U,D,E}$ are the $3\times 3$ Yukawa matrices for the quarks
and charged leptons. The soft breaking parameters
$A_i$ and $B$ are dimensionless numbers of order unity.

The $b\rightarrow s\gamma$ and $b\rightarrow s\gamma\gamma$
transitions can proceed in the SUSY model
 via five different intermediate particles exchanges:
\begin{enumerate}
\item Charged gauge bosons $(W^-)\ \ \ +\ \ \ $ up-quarks
\item Charged Higgs bosons $(H^-)\ \ \ +\ \ \ $ up-quarks
\item Charginos $(\chi^-)\ \ \ +\ \ \ $ up-squarks
\item Gluinos $({\widetilde g})\ \ \ +\ \ \ $ down-squarks
\item Neutralinos ($\chi^0)\ \ \ +\ \ \ $ down-squarks
\end{enumerate}

The total amplitude is the sum of all these
contributions. The complete analytic expressions for the
various components are found in ref. \cite{BBMR}.

An effective $b-s$ flavour changing transition induced by $W^-$ exchange
is the only way through which the decays proceed in
the SM. A two-Higgs doublet extension of the SM
would include the first two contributions,
while the last three are genuinely supersymmetric in nature.

Gluinos and neutralinos can mediate flavour changing interactions
only via renormalization effects which are crucially dependent on the
detailed structure of the model. Their consideration is beyond the scope
of the present work and our results do not presently
justify a more detailed analysis.
We shall discuss the features of
the inclusion of the first three contributions in the matching
of the Wilson coefficients.

The supersymmetric Wilson coefficients are then given by
\begin{equation}
C_{7,8}^{SUSY} (m_W) =
C_{7,8}^{SM}(m_W) + C_{7,8}^H (m_W) + C_{7,8}^\chi (m_W)\ ,
\end{equation}
while at the LO the matching conditions in
eqs.~(\ref{c1-6})--(\ref{c2}) remain unaffected.

From the results of ref.~\cite{BBMR} and comparing with
eq.~(\ref{heff}) we obtain the following contributions
to the $C_{7,8}(m_W)$ coefficients:
\begin{equation}
C_{7,8}^H (m_W) =\frac{1}{2}\frac{m_t^2}{m_H^2}\left[
\frac{1}{\tan^2 \beta}
f^{(1)}_{7,8} \left( \frac{m_t^2}{m_H^2}\right) +
f^{(2)}_{7,8} \left( \frac{m_t^2}{m_H^2}\right) \right]\ ,
\label{c78h}
\end{equation}
induced by charged Higgs exchange and
\begin{eqnarray}
C_{7,8}^\chi (m_W) &=&
   - {1 \over V_{ts}^* V_{tb}} \
   \sum_ {j=1} ^2 \sum_ {k=1} ^6 \
   {m_W^2 \over \tilde m_{\chi_j}^2}
   \left[
   (G_{UL}^{jkb} - H_{UR}^{jkb})
   (G_{UL}^{*jks} - H_{UR}^{*jks})
   f_{7,8}^{(1)}\left(\frac{\tilde m_{u_k}^2}{\tilde m_{\chi_j}^2}\right)
   \right.                                    \nonumber    \\
 & & - \left.
    H_{UL}^{jkb} (G_{UL}^{*jks} - H_{UR}^{*jks}) \
   {\tilde m_{\chi_j} \over m_b} \
   f_{7,8}^{(3)}\left(\frac{\tilde m_{u_k}^2}{\tilde m_{\chi_j}^2}\right)
   \right]
\label{c78ch}
\end{eqnarray}
induced by chargino exchange.
We have found convenient for the present discussion
to introduce the functions $f_{7,8}^{(n)}$ according to the notation
of ref.~\cite{barbieri}
\begin{eqnarray}
f^{(1)}_7 (x) &=& \frac{(7-5x-8x^2)}{36(x-1)^3}+
\frac{x(3x-2)}{6(x-1)^4}\log x
\label{f1}\\
f^{(2)}_7 (x) &=& \frac{(3-5x)}{6(x-1)^2}+
\frac{(3x-2)}{3(x-1)^3}\log x \\
f^{(3)}_7 (x) &=& (1-x) f^{(1)}_7 (x) -
\frac{x}{2}f^{(2)}_7 (x) -\frac{23}{36} \\
f^{(1)}_8 (x) &=& \frac{(2+5x-x^2)}{12(x-1)^3}-
\frac{x}{2(x-1)^4}\log x \\
f^{(2)}_8 (x) &=& \frac{(3-x)}{2(x-1)^2}-
\frac{1}{(x-1)^3}\log x \\
f^{(3)}_8 (x) &=& (1-x) f^{(1)}_8 (x) -
\frac{x}{2}f^{(2)}_8 (x) -\frac{1}{3}\ .
\label{f3}
\end{eqnarray}
These functions have simple and obvious relations with the
functions $F_n (x)$ defined originally in ref.~\cite{BBMR}, to which we
refer the reader for all details.

In eq.~(\ref{c78ch}) $j=1,2$ is the label of the chargino
mass eigenstates and $k=1,...,6$ is the analogous
label for the up-squarks;
the matricial couplings $G_{UL}$
arise from charged gaugino-squark-quark vertices,
whereas $H_{UL}$ and $H_{UR}$ are related to the charged
higgsino-squark-quark vertices.
These couplings contain among else
the unitary rotations $U$ and $V$
which diagonalize the chargino mass matrix
\begin{equation}
U^\ast \pmatrix{M_2 & m_W\sqrt{2} \sin \beta \cr
                m_W\sqrt{2} \cos \beta & -\mu \cr}
V^{-1} =\pmatrix{\tilde{m}_{\chi_1} & 0 \cr
                 0 & \tilde{m}_{\chi_2} \cr}\ ,
\label{chmatrix}
\end{equation}
where $M_2$ is the weak gaugino mass and $\mu$ the Higgs mixing parameter.
The sign of the $\mu$ entry is defined accordingly to the Feynman rules
used in obtaining the above results (see the comment following eq. (15)
in ref.~\cite{bevi}).

Due to the relevance of the chargino amplitude
for the present discussion, it is worth
trying to have a better understanding
of the nature of the features exhibited
by this amplitude.

An explicit $\tan\beta$ dependence is found
in $H_{UL}$ and $H_{UR}$ where
quark Yukawa couplings are present; more precisely,
$H_{UL}$ is proportional to the down-quark Yukawa coupling,
which grows with $\tan\beta$ as $1/\cos\beta$,
whereas $H_{UR}$ contains the up-quark Yukawa coupling,
that approaches in the large $\tan\beta$ limit
a constant value ($\propto 1/\sin\beta$).
It is in fact the contribution in the second line of eq.~(\ref{c78ch})
that determines the behaviour of the amplitude in the
large $\tan\beta$ regime.
Detailed studies of this feature of the chargino amplitude are available
in the literature~\cite{largebeta}. 

In order to allow for an analytic  and more transparent discussion
of the chargino component we resort to simplified assumptions on the
squark mass spectrum which reproduce with good approximation the
global features of the model. In this we follow closely the analysis
of ref.~\cite{barbieri}.
We assume that all squarks, other than the two scalar
partners of the top quark, are
degenerate with the soft breaking mass $\tilde{m}$.
The remnant $2\times 2$ top squark mass
matrix is diagonalized by an orthogonal matrix $T$ such that:
\begin{equation}
T\pmatrix{\tilde{m}^2+m_t^2 & (A_t\tilde{m} + \mu/\tan\beta)m_t \cr
          (A_t\tilde{m} + \mu/\tan\beta) m_t & \tilde{m}^2+m_t^2 \cr}T^{-1}=
\pmatrix{\tilde{m}^2_{t_1}&0\cr 0& \tilde{m}^2_{t_2}\cr},
\end{equation}
where $A_t$ is the supersymmetry-breaking trilinear coupling.
The sign of the $\mu$ term is consistent with eq.~(\ref{chmatrix}).

With these assumptions eq.~(\ref{c78ch}) can be written as
\begin{eqnarray}
C_{7,8}^\chi (m_W) &=& \sum_{j=1}^{2} \left\{
\frac{m_W^2}{\tilde{m}_{\chi_j}^2}\left[ |V_{j1}|^2
f^{(1)}_{7,8} \left( \frac{\tilde{m}^2}{\tilde{m}_{\chi_j}^2}\right)
\right. \right. \nonumber \\
&&-\left. \sum_{k=1}^2 \left| V_{j1}T_{k1}-V_{j2}T_{k2}\frac{m_t}{\sqrt{2}
m_W \sin \beta} \right|^2
f^{(1)}_{7,8} \left( \frac{\tilde{m}_{t_k}^2}
{\tilde{m}_{\chi_j}^2}\right) \right] \nonumber \\
&&-\frac{U_{j2}}{\sqrt{2} \cos \beta}
\frac{m_W}{\tilde{m}_{\chi_j}}\left[ V_{j1}
f^{(3)}_{7,8} \left( \frac{\tilde{m}^2}{\tilde{m}_{\chi_j}^2}\right)
\right. \nonumber \\
&&-\left. \left.
\sum_{k=1}^2 \left( V_{j1}T_{k1}-V_{j2}T_{k2}\frac{m_t}{\sqrt{2}
m_W \sin \beta} \right) T_{k1}
f^{(3)}_{7,8} \left( \frac{\tilde{m}_{t_k}^2}
{\tilde{m}_{\chi_j}^2}\right) \right]  \right\}\ ,
\label{c78chb}
\end{eqnarray}

\subsection{\bf Four exemplifying cases }

We are now ready to investigate the effects of the SUSY matchings
on the $b\to s\gamma\gamma$ transition.
We will show our results by plotting the ratios of SUSY versus SM
decay rates for central values of the SM input parameters
while varying the unknown SUSY parameters.
We investigate four limiting cases which span the global features
of the new amplitudes.


First, we take $M_2=\mu=A_t=0$ and fix $\tan\beta=1$ (case 1).
In this approximation we have
\begin{equation}
U=\frac{1}{\sqrt{2}}\pmatrix{1&1\cr -1&1},~~~
V=\frac{1}{\sqrt{2}}\pmatrix{1&1\cr 1&-1},~~~
\tilde{m}_{\chi_{1,2}}=m_W,
\end{equation}
\begin{equation}
T=\pmatrix{1&0\cr 0&1},~~~ \tilde{m}^2_{t_{1,2}}=\tilde{m}^2+m_t^2\ .
\end{equation}
The Wilson coefficients $C_{7,8}^\chi$ can be simply written as:
\begin{equation}
C_{7,8}^\chi (m_W) =
z\left[ f^{(1)}_{7,8}(z) +\frac{1}{2} f^{(2)}_{7,8}(z)\right]
-(2x+z)f^{(1)}_{7,8}(x+z) -\frac{x+z}{2} f^{(2)}_{7,8}(x+z),
\end{equation}
where
\begin{equation}
x= \frac{m_t^2}{m_W^2},~~~
z= \frac{\tilde{m}^2}{m_W^2}.
\end{equation}

As remarked in ref.~\cite{barbieri} $C_{7,8}^{SUSY}(m_W)$ shows an exact
cancellation in the supersymmetric limit, $z\to 0$ and $m_H\to m_W$.
This is a consequence of the fact that any magnetic moment transition
vanishes in exact supersymmetry~\cite{ferrara}.
Therefore non-vanishing contributions to the  $C_{7,8}^{SUSY}$
coefficients arise due to the presence of the soft breaking terms.

We define
\begin{equation}
R_{\gamma\gamma} = \frac{BR(B_s\to\gamma\gamma)_{SUSY}}{
BR(B_s\to\gamma\gamma)_{SM}}\ ,
\label{rff}
\end{equation}
and
\begin{equation}
R_{\gamma} = \frac{BR(B\to X_s\gamma)_{SUSY}}{
BR(B\to X_s\gamma)_{SM}}\ ,
\label{rf}
\end{equation}
where the SM decay rates are those given in
eqs.~(\ref{brsfsm})--(\ref{brffsm}),
obtained using the central values of the SM input parameters.

\epsfxsize=8cm
\centerline{\epsfbox{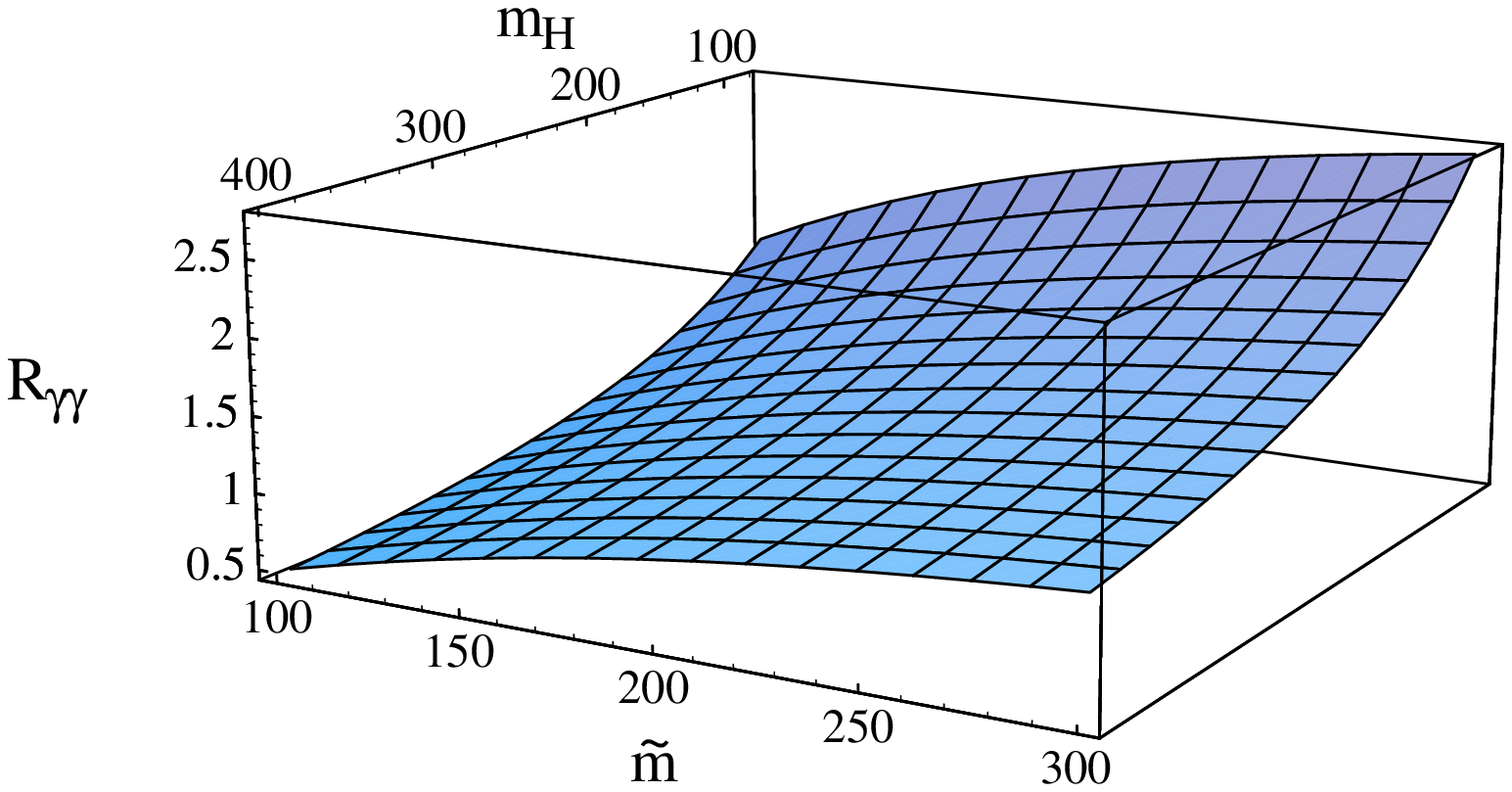}}
\label{mfig3}
{Fig. 3.
Case 1. $R_{\gamma\gamma}$
as a function of $\tilde{m}$ and $m_H$ (GeV),
for degenerate chargino masses $\tilde m_{\chi_{1,2}}=m_W$.
}

\epsfxsize=8cm
\centerline{\epsfbox{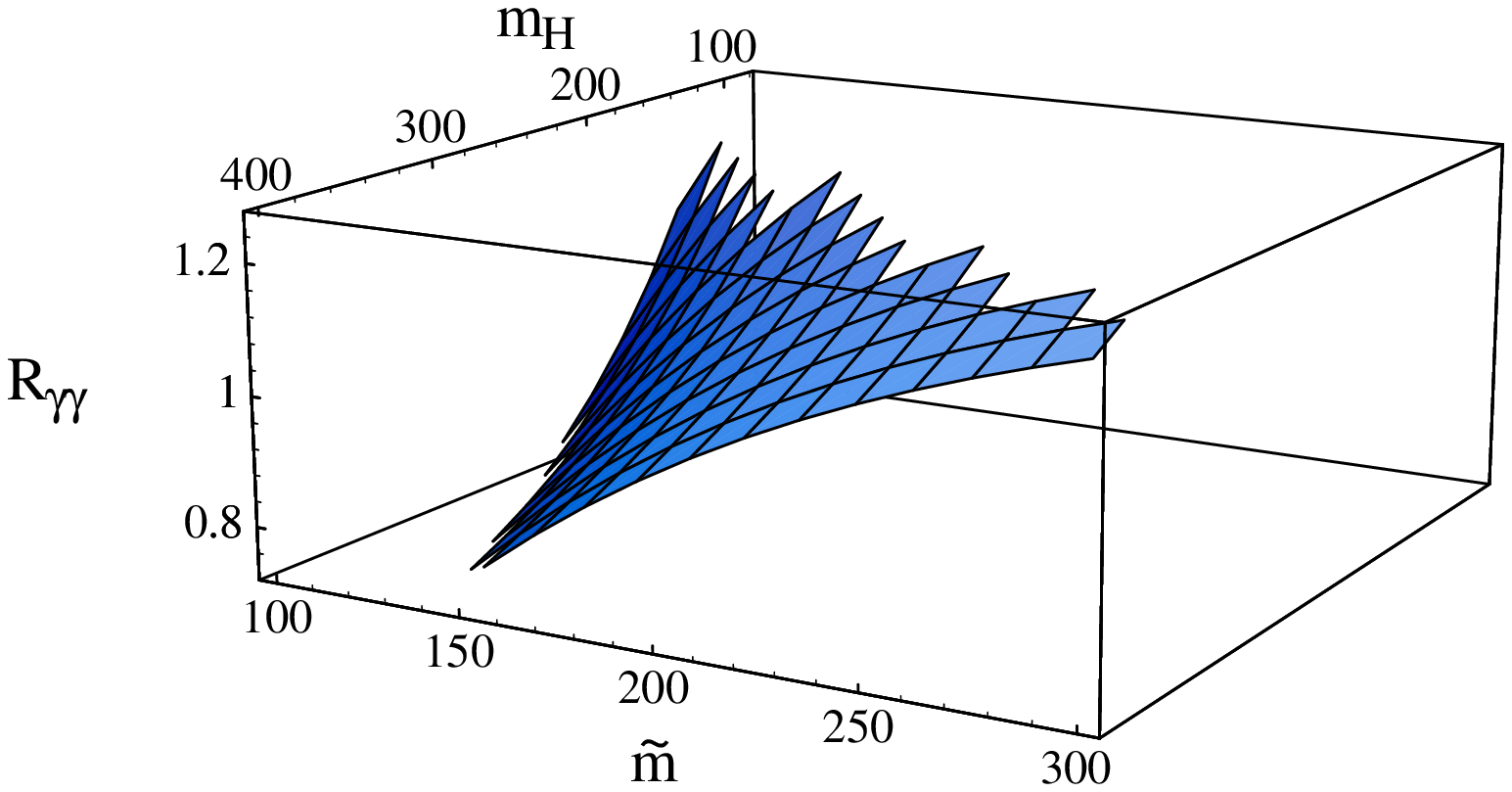}}
\label{mfig4}
{Fig. 4.
Case 1. The allowed range for $R_{\gamma\gamma}$ is shown
as a function of $\tilde{m}$ and $m_H$ (GeV),
by constraining
the $BR(B\to X_s \gamma)_{SUSY}$ to vary
within a $\pm 30 \%$ from its SM value.
}

In Fig. 3 we show $R_{\gamma\gamma}$ as a function of the charged Higgs
mass and the scalar soft breaking mass in a few hundred GeV range.
In Fig. 4 the same range is spanned assuming the constraint
\begin{equation}
0.7 < R_\gamma < 1.3\ .
\label{constraint}
\end{equation}
We see that $R_{\gamma\gamma}$ as well is bound to vary in approximately
the same range. The study of the ratio $R_{\gamma\gamma}/R_\gamma$
in the same region
shows deviations of at most $\pm 4\%$ from unity, which shows the
strong correlation between the two decays.

\bigskip


By releasing the constraint $A_t = 0$, while holding $M=\mu=0$
and $\tan\beta=1$,
we allow for a mass splitting of the stop eigenstates (case 2).
This corresponds to having
\begin{equation}
T=\frac{1}{\sqrt{2}}\pmatrix{1&1\cr -1&1},~~~
\tilde{m}^2_{t_{1,2}}=\tilde{m}^2+m_t^2\pm A_t\tilde{m} m_t\ .
\label{stopmass}
\end{equation}
The chargino contribution to $C_{7,8}$ becomes:
\begin{equation}
C_{7,8}^\chi (m_W) =zf^{(1)}_{7,8} (z)+
\frac{z}{2}f^{(2)}_{7,8} (z)- \sum_{k=1}^2 \left[
\frac{x+w_k}{2}f^{(1)}_{7,8} (w_k) +
\frac{w_k}{4}f^{(2)}_{7,8} (w_k)\right] .
\end{equation}
where $w_{1,2}=x+z \pm A_t\tilde{m} m_t/m_W^2$.

%
\epsfxsize=8cm
\centerline{\epsfbox{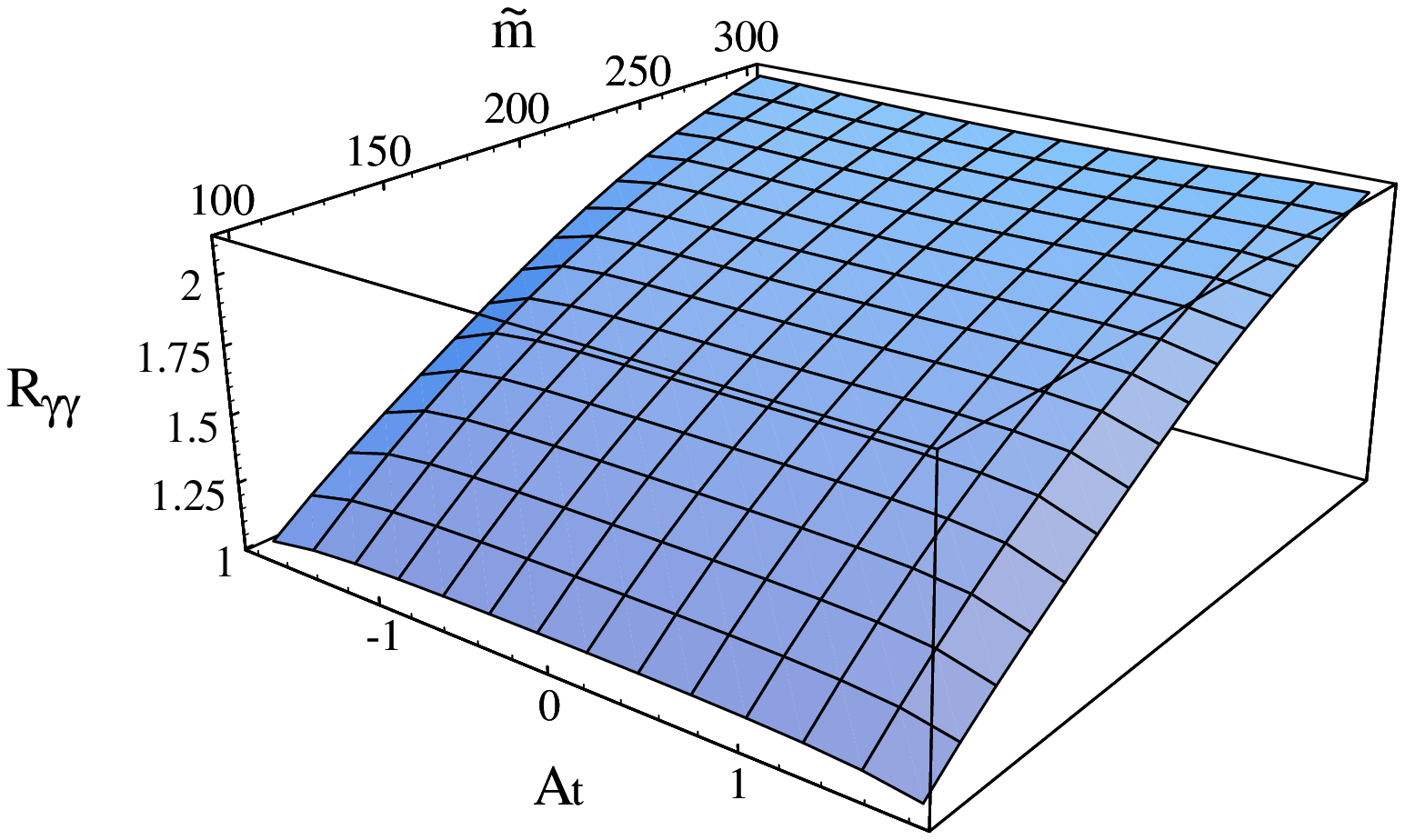}}
\label{mfig5}
{Fig. 5.
Case 2.
$R_{\gamma\gamma}$ is shown as a function of $\tilde{m}$ (GeV) and
$A_{t}$ for $m_H=150$ GeV.
}

In Fig. 5 we plot $R_{\gamma\gamma}$ as a function of $A_t$
and $\tilde m$ for fixed $m_H=150$ GeV, under the requirement
that the lightest stop mass is always above 45 GeV.
As we see, releasing the stop squark degeneracy while keeping
charginos degenerate does
not sizeably modify the features shown in case 1,
and the same conclusions apply.

\bigskip


Next we consider $M=\mu =A_t=0$, and arbitrary $\tan \beta$
(case 3).
The chargino component reduces to:
\begin{eqnarray}
C_{7,8}^\chi(m_W) &=&
-\frac{x+z}{4\cos^4 \beta}\left[
f^{(1)}_{7,8} \left( \frac{x+z}{2\cos^2\beta}\right) +
\frac{1}{2}
f^{(2)}_{7,8} \left( \frac{x+z}{2\cos^2\beta}\right) \right]
-\frac{x}{4\sin^4\beta}
f^{(1)}_{7,8} \left( \frac{x+z}{2\sin^2\beta}\right)
\nonumber \\
&& + \frac{z}{4\cos^4 \beta}\left[
f^{(1)}_{7,8} \left( \frac{z}{2\cos^2\beta}\right) +
\frac{1}{2}
f^{(2)}_{7,8} \left( \frac{z}{2\cos^2\beta}\right) \right]\ .
\label{case3}
\end{eqnarray}

In this case, the chargino degeneracy is lifted, while keeping
the degeneracy in the squark sector.
The chargino contribution becomes dependent on $\tan\beta$.
On the other hand, as can be verified by means of
eqs. (\ref{f1})--(\ref{f3}),
the $\tan\beta$ dependence of eq.~(\ref{case3}) in the large $\tan\beta$
limit is only logarithmic. As we will see later, for
the SUSY amplitude to exhibit a stronger $\tan\beta$
dependence $A_t\ne 0$ is required as well.

\epsfxsize=8cm
\centerline{\epsfbox{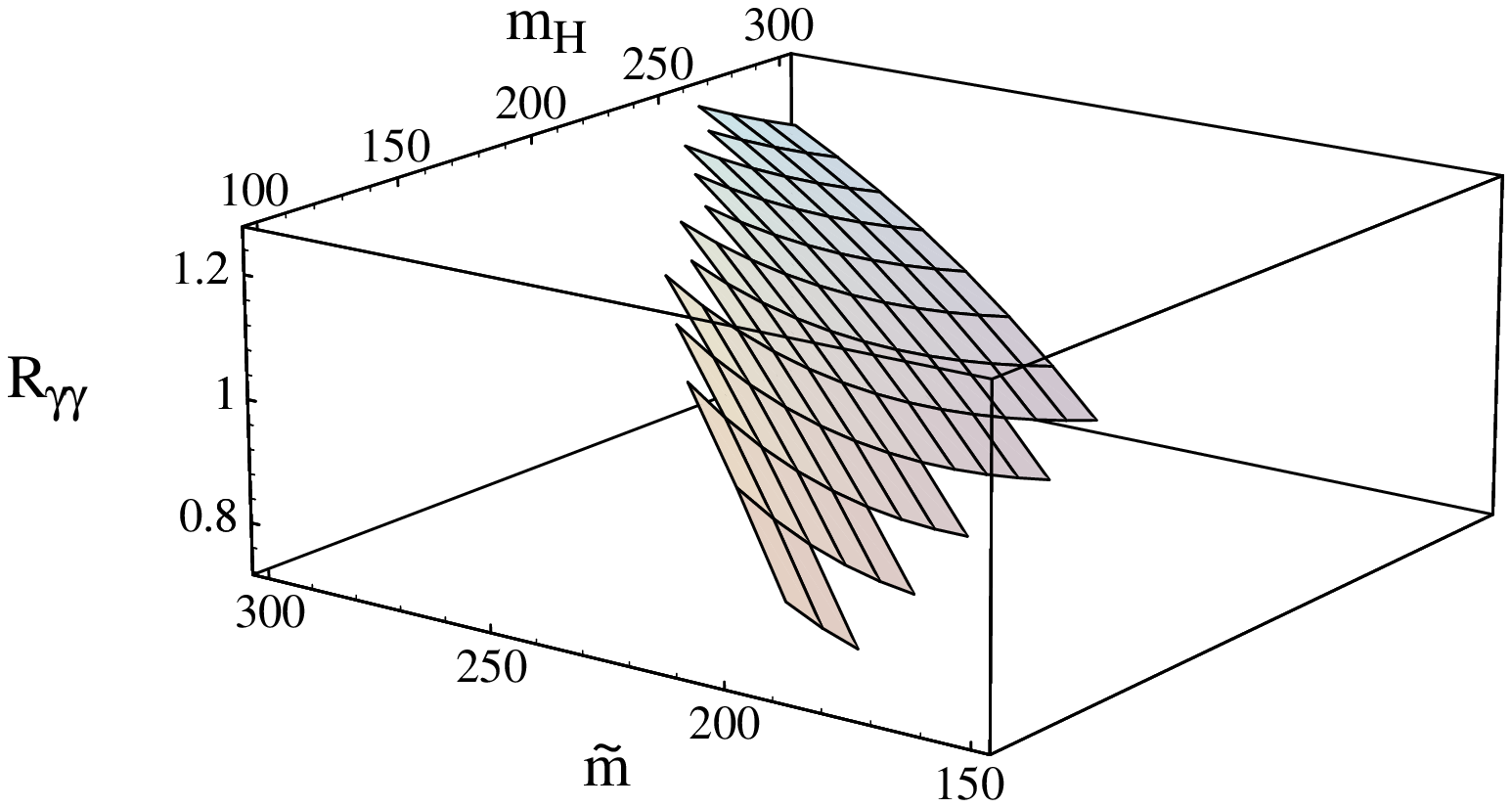}}
\label{mfig6}
{Fig. 6.
Case 3. The allowed range for
$R_{\gamma\gamma}$ is shown as a
function of $\tilde{m}$ and $m_H$ (GeV)
for $\tan \beta=10$, imposing the constraint
of eq.~(\ref{constraint}) on $BR(B\to X_s\gamma)_{SUSY}$.
}

\epsfxsize=8cm
\centerline{\epsfbox{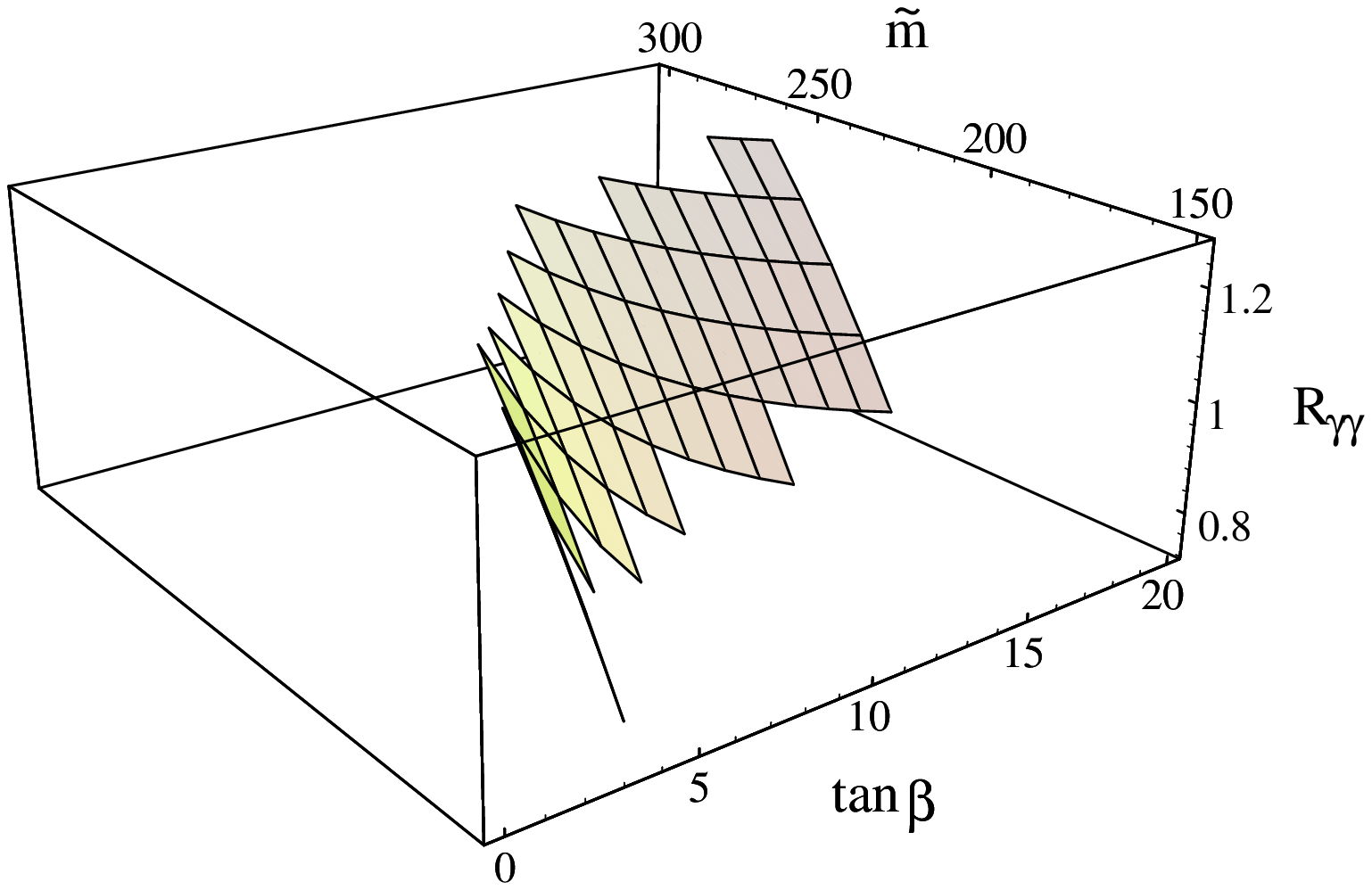}}
\label{mfig7}
{Fig. 7.
Case 3. The allowed range for
$R_{\gamma\gamma}$ is shown as a
function of $\tilde{m}$ (GeV) and $\tan\beta$
for $m_H=150$ GeV, imposing the constraint
of eq.~(\ref{constraint}) on $BR(B\to X_s\gamma)_{SUSY}$.
}
\bigskip

In Figs. 6 and 7 we show as a function of different SUSY parameters
the allowed range for $R_{\gamma\gamma}$ once the constraint
on $BR(B\to X_s\gamma)_{SUSY}$ in eq.~(\ref{constraint}) is imposed.
$R_{\gamma\gamma}$ is always bound to vary within $\pm 30$\%
from its SM expectation with high correlation to $R_{\gamma}$.


Finally we consider the case
for which $M_2$, $\mu$  and $A_t$ are different
from zero and $\tan\beta \gg 1$ (case 4).
In fact, the part of the chargino
contribution which leads the large $\tan\beta$ behaviour
vanishes when either squarks or charginos are degenerate, as one can verify
from the simplified form of eq.~(\ref{c78chb}).

\epsfxsize=8cm
\centerline{\epsfbox{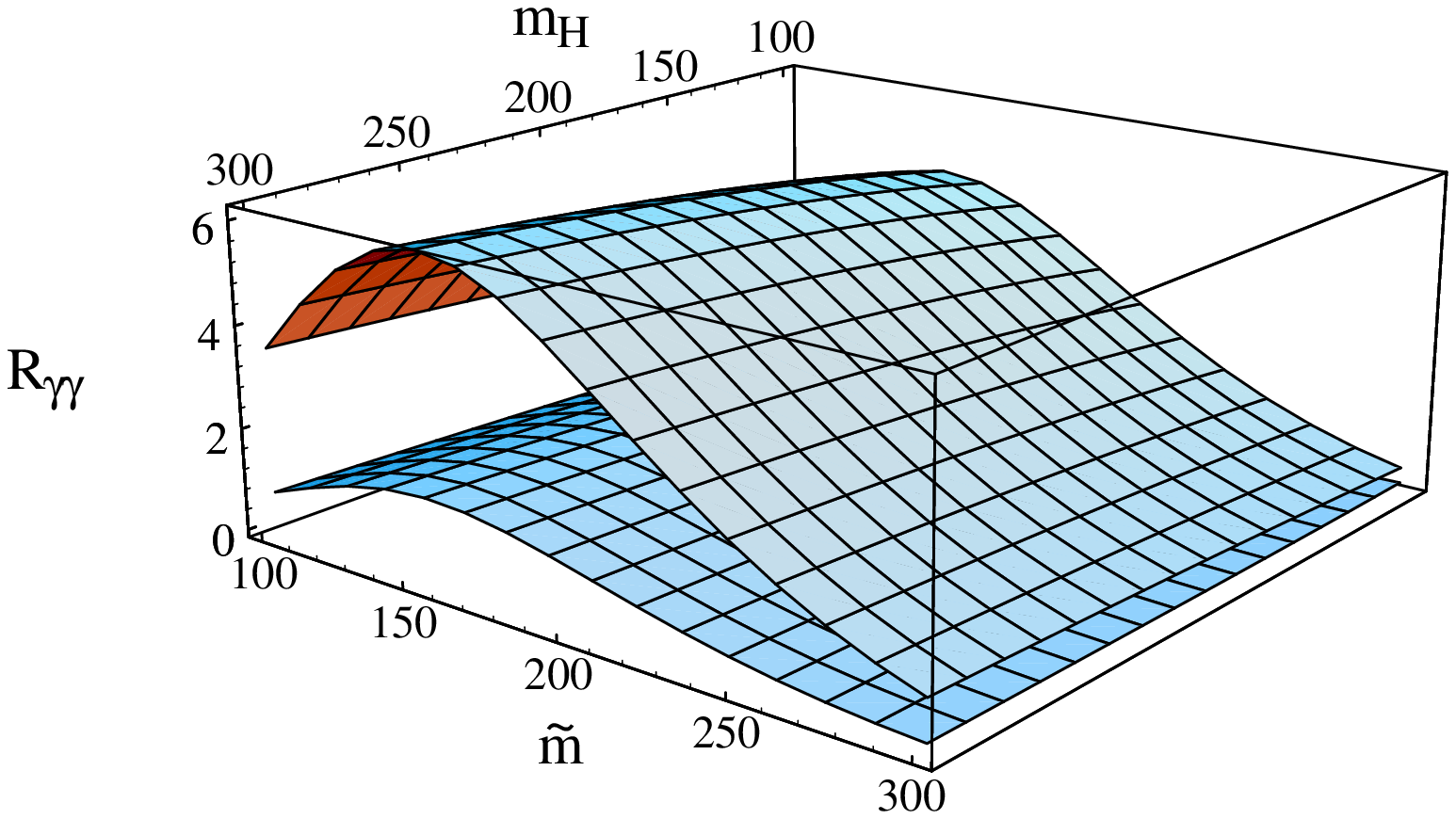}}
\label{mfig8}
{Fig. 8.
Case 4.
The potential enhancement of
$R_{\gamma\gamma}$ is shown as a
function of $\tilde{m}$ and $m_H$ (GeV)
for $\tan \beta=15$ (upper surface) and 10 (lower surface).
}
\bigskip

An analytic approximation of the $H_{UL}H^*_{UR}$ component
of the chargino amplitude
can be derived which shows explicitly
its interesting features~\cite{bevi}.
We assume the chargino mass matrix
in eq.~(\ref{chmatrix}) to be approximately diagonal:
\begin{equation}
M_{\chi}\approx {\rm diag}(M_2,-\mu)
\label{diagonal m-chi}
\end{equation}
This approximation holds effectively when~\cite{bevi}
$|M_2^2 - \mu^2| = $
$O[{\rm max}(M_2^2,\ \mu^2)] \gg m_W^2$ and
$(M_2^2,\ \mu^2) \ \raisebox{-.4ex}{\rlap{$\sim$}} \raisebox{.4ex}{$>$}\ m_W^2$.
It is important to notice that
these requirements, and therefore the approximation
of eq.~(\ref{diagonal m-chi}) is consistent with one of the
eigenvalues, say $|\mu|$,
being of the order of $m_W$, while the other $|M_2|$ is much heavier.
Thehe chargino mass matrix already
being diagonal, the approximate mass eigenvalues are simply
given by the absolute values
of the parameters $M_2$ and $\mu$, and the
two unitary rotations which ``diagonalize'' the
chargino mass matrix can be written as:
\begin{equation}
\begin{array}{ccl}
U &\approx & {\rm diag (sign}[M_2],-{\rm sign}[\mu])\ , \\
V &\approx & {\bf 1}
\end{array}
\label{approx decomposition}
\end{equation}
In this approximation, the matrix $T$ and the stop mass eigenstates are
given by eq.~(\ref{stopmass}).

Using eq.~(\ref{stopmass}) and eqs.~(\ref{diagonal
m-chi})--(\ref{approx decomposition})
we obtain a simple expression for the part of the chargino component
relevant for large $\tan\beta$:
\begin{equation}
C_{7,8}^\chi (m_W) \approx
\frac{1}{2 \sin 2\beta}
\frac{m_t}{\mu} \left[f_{7,8}^{(3)}\left(\frac{\tilde m^2_{t_1}}{\mu^2} \right)
- f_{7,8}^{(3)}\left(\frac{\tilde m^2_{t_2}}{\mu^2} \right)
\right]
\label{domin ampl}
\end{equation}
where
$\tilde m_{\chi_2}=|\mu|$ is the lightest chargino eigenvalue.
Notice that the amplitude in eq.~(\ref{domin ampl}) depends on the signs
of both $\mu$ and the trilinear soft breaking parameter $A_t$ (changing
the sign of the latter amounts to interchanging the two stop mass eigenvalues).
One also verifies that the amplitude
vanishes for either $\mu= 0$ or $A_t=0$ as it should.

As already mentioned, at variance with the case 3, the leading
behaviour of the chargino amplitude is linear with $\tan\beta$.
This may in general be the source of large deviations
of the SUSY rates from the corresponding SM expectations.
In Fig. 8 we show the ratio of the SUSY to SM branching ratio
for $B_s\to\gamma\gamma$ as a function of $m_H$ and
$\tilde m$ for $\tan\beta=10$ and 15. In the example shown we have
chosen $\mu=m_W$ and $A_t=1.5$.
We see the potential large enhancements which arise for large
$\tan\beta$ from this component of the chargino amplitude.
On the other hand, imposing the constraint in eq.~(\ref{constraint})
allows only those regions of the lower surface
for which $R_{\gamma\gamma}$ varies approximately in the range
$0.7-1.3$.

Globally, in the tested region of parameters
the deviations of the $B_s\to\gamma\gamma$
decay rate from the SM expectations are confined to be
within 10\% from the corresponding deviations for the
$B\to X_s\gamma$ decay.

At the next-to-leading order one may try to devise models that
enhance the matchings of the $O_{3-6}$ penguin operators (which are
vanishing at the LO)
keeping the $O_{7,8}$ Wilson coefficients ``under control''.
As unlikely as this may be, numerically it is anyhow difficult
to expect drastic deviations from the SM predictions, due to the
subleading role of the $O_{3-6}$ operators (analogous
considerations apply to the electroweak penguin operators,
which we have neglected in the LO analysis).

We conclude that in order to disentangle new physics effects from a
comparison of the two $b\to s$ radiative decays a precision
below 10\% is required both on the theoretical and experimental sides.
Due to the smallness of the two-photon rates and to the theoretical
uncertainties related to long-distance physics it shows as a truly
challenging task.

\acknowledgments{
S. Bertolini thanks the Physics Department at the University of Oslo 
for the financial support and hospitality during the completion
of this work.
J. Matias acknowledges financial support from Ministerio de Educacion
y Ciencia.
The authors thank N. Di Bartolomeo for contributing to
preliminary discussions.
}

%
%
\vspace{1cm}
%
%
\renewcommand{\baselinestretch}{1}

\begin{table}
        \begin{center}
      \begin{tabular}{l l}
     \, $\alpha_{s}(m_Z)$  \, 	& \, $0.118$ \,            \\
     \, $\alpha_{em}$		   \,   	& \,  1/129   \,        \\
     \, $m_{Z}$          \,  		& \, $91.19$ GeV \, \\
     \, $m_{W}$          \,   		& \, $80.33$ GeV \, \\
     \, $m_{t}$          \,   		& \, $175$ GeV  \, \\
     \, $m_b$             \,      	& \, $4.8$ GeV \, \\
     \, $m_c$              \,     	& \, $1.4$ GeV  \, \\
     \, $m_s$               \,    	& \, $0.150$ GeV \, \\
     \, $|V_{ts}^*V_{tb}|/|V_{cb}|$ \, 	& \, $0.976$  \, \\
     \,  	$|V_{ts}^*V_{tb}|$		 \, & \, $4\times 10^{-2}$  \, \\
    \,    $m_{B_s}$           \,  		& \, $5.37$ GeV \, \\
     \,   $f_{B_s}$           \,  		& \, $0.2$ GeV \,  \\
    \,    $\Gamma(B_s)$			 \, & \, $4.09 \times 10^{-13}$ GeV \,
\\
    \,    $BR(B\to X_c e\bar\nu_e)$	 \, & \, $10.4 \times
10^{-2}$ \, \\
        \end{tabular}
        \end{center}
\caption{Values of the input parameters used in the numerical
          calculations.}
\label{inputs}
\end{table}

\end{document}